Master's Thesis of Engineering

# Clustering Approaches for Global Minimum Variance Portfolio

February 2020

Graduate School of Convergence Science and Technology

Seoul National University

Jinwoo Park

# Abstract

To earn higher return, one must bear higher risk, as risk and return are trade-off. However, a portfolio of well diversified assets allows investors to earn the same return at the expense of less amount of risk. It is because price of various assets can move in the opposite way, offsetting each other and thus, resulting in a more stable portfolio with less volatile returns.

Due to recent stock market crashes, portfolio optimization methods which focus on investing safely receive attention from various investors. Global minimum variance portfolio (GMVP) is an investment strategy designed to carry as little variance, which is considered risk in finance, as possible. The only input to attain the portfolio weights of GMVP is the covariance matrix of asset returns. Since the population covariance matrix is not known, investors use historical data to estimate covariance matrix. Even though sample covariance matrix is an unbiased estimator of the population covariance matrix, it includes a great amount of estimation error especially when the number of observed data is not much bigger than number of assets. It is due to the fact that bigger number of variance and covariance parameters need to be estimated and the covariance matrix approaches singularity.

Clustering stocks is proposed to decrease the estimation error contained in the covariance matrix and inverse of it by reducing the number of features. As it is difficult to estimate the covariance matrix with high dimensionality, we can perform portfolio optimization in each cluster first, and then perform portfolio optimization once again between clusters. In this sense, clustering approach to portfolio optimization is 'divide and conquer'.

The motivation of this dissertation is that the estimation error can still remain high even after clustering, if a large amount of stocks is clustered together in a single group. This research, thus, proposes to utilize a bounded clustering method in order to limit the maximum cluster size. The result of one example experiment shows that not only the gap between in-sample volatility and out-of-sample volatility decreases, but also the out-of-sample volatility decreases. Moreover, other risk measures such as downside standard deviation and maximum drawdown tends to diminish. By limiting the maximum cluster size while clustering stocks, investors can better predict the out-of-sample risk based on in-sample counterpart and expect smaller out-of-sample portfolio risks.

There are three academic contributions of this dissertation. Firstly, this research shows that when we utilize clustering approach to portfolio optimization, clustering quality and estimation error are trade-off and maximum clustering size influence both. For example, as maximum clustering size increases, clustering quality improves, whereas the estimation error becomes larger, and vice versa. Secondly, we illustrate that bounded clustering approach is needed to find the best maximum clustering size to find the compromise between the clustering quality and estimation error to achieve the best portfolio performance. Thirdly, portfolio performance improvement from scaling data and applying dimensionality reduction results from taking care of estimation error while clustering stocks.

**student number**: 2018-28965

# Table of Contents

# List of Tables

# List of Figures

# 1. Introduction

This chapter provides the overview of our research. Section 1.1 summarizes background and motivation of our research and section 1.2 presents the goal of the research. Section 1.3 briefly introduces the overview of the experiment, whereas section 1.4 touches upon the academic contribution of the research.

## 1.1 Background and Motivation

**Portfolio optimization**

As the phrase 'high risk, high return' implies, if one targets to earn higher return, he must do so at the expense of higher risk. To put it differently, there is a trade-off between risk and return [1]. By making investment in wide variety of assets showing different return movements, however, lets us earn the same expected return with less amount of expected risk. It is because, for example, the negative return of one asset can be offset by the positive return of another asset, resulting in a stable portfolio as a whole. As deviation from expected return is considered risk, having a portfolio of diversified assets can reduce risk of investment. This risk-reduction available from portfolio investment is now a commonly accepted investment strategy [2]. The rest of the dissertation focuses on investing in equity (i.e. stocks).

To have a portfolio of diversified assets and benefit from diversification, two questions need to be answered: firstly, how to pick the best suitable stocks to put in a portfolio and secondly, how to decide upon the weights of each chosen stock. Portfolio optimization is a mathematical attempt to answer both of these two questions. It chooses a group of stocks and determines weights of each stock in a portfolio such that it best suits the given objective such as minimizing a variance of portfolio returns or maximizing a risk-reward ratio. There have been many portfolio optimizations developed since Harry Markowitz introduced Modern Portfolio Theory (MPT) in1952 [3].

**Global Minimum Variance Portfolio**

Due to the recent stock market crashes such as the stock market selloff in 2016 and world-wide stock market downturn in 2018, investors have increasing appreciation for risk management in their equity investment. The global minimum variance portfolio (GMVP) is the well-known investment strategy that addresses these risk-sensitive investors because it targets to carry as little volatility as possible. The portfolio is located at the outmost left point on the efficient frontier. The efficient frontier is a group of the portfolios with possibly lowest variance given certain target returns.

An investor cannot hold a portfolio of risky assets with the lower risk than the GMVP, if the input for the algorithm to construct the GMVP is known. The only input needed for GMVP is a covariance matrix of asset returns that are being considered for investment. Output of GMVP is a vector of portfolio weights, which tells investors how much of wealth should be invested in each stock for achieving the portfolio with the least amount of variance. Since the population covariance matrix is unknown in practice, it needs to be estimated from historical data. Sample covariance matrix is frequently used as it is an unbiased estimator of population covariance matrix.

**Estimation error in the sample covariance matrix**

The problem of the sample covariance matrix arises when the number of features (number of stocks being considered for creating a portfolio) is bigger than or as big as the number of observations (daily return, weekly return, etc.). As the number of assets increase, not only more variance and covariance parameters have to be estimated, but also the sample covariance matrix approaches singularity. Both of these two problems result in estimation error in the covariance matrix and inverse of covariance matrix, which causes the portfolio weights obtained from GMVP formula to become unstable and sensitive to small input changes.

As a consequence of using poorly estimated portfolio weights, the out-of-sample portfolio variance will be much higher than the in-sample counterpart [4]. The difference between the out-of-sample performance and in-sample counterpart is due to estimation error contained in covariance matrix and inverse of covariance matrix. This poor out-of-sample performance will frustrate investors expecting investment with small amount of variance and it might underperform even randomly selected portfolios [5]. It is also confirmed by another research that the portfolio made with equal weights in stocks may perform better than GMVP [6].

**Clustering approach to GMVP**

There have been many attempts to decrease the estimation error, using more structured estimator than the sample covariance matrix to estimate the population covariance matrix. One of these attempts is to decrease the number of features by clustering similar stocks and considering them as a single stock. By doing so, the number of variance and covariance parameters to be estimated decreases and the covariance matrix gets farther away from singularity.

Rather than coming up with the covariance of all the stocks and optimizing a portfolio all at once, this clustering approach to portfolio optimization follows **'divide and conquer'**. It divides the stocks into clusters and performs a portfolio optimization within each cluster first. With the result of portfolio optimization attained from each cluster, the portfolio optimization is performed once again to come up with the cluster weights. By multiplying the cluster weights and stock weights within each cluster, we can calculate portfolio weights. This approach fulfils the goal of computing the covariance matrix of assets in two steps.

Several clustering approaches have been suggested so far; some utilize non-price information such as industry sectors and other clustering approaches employ price. The drawback of using non-price information in clustering stocks is that stock price often diverges from the direction of non-price information. This divergence is of a critical issue because if stocks showing different price movements are clustered together, cluster quality degenerates, resulting in higher correlation between stocks and poor portfolio performance.

While taking care of estimation error in GMVP, we also need to pay attention to the clustering quality because portfolio performance is affected not only by estimation error but also correlation between clusters. Clustering quality is great when both intra-cluster similarity and inter-cluster dissimilarity is high. The better clustering quality, the less correlation between clusters is. In portfolio optimization, the less correlation between the assets leads to the smaller standard deviation of portfolio returns [7]. As such, clustering based on daily return of stocks seems appropriate.

The problem of clustering stocks based on stock price is that high estimation error may still remain If many stocks are clustered under a single cluster. In some trading days, most stocks move towards the same direction with relatively significantly than other trading days. For illustration, when market declines, price of most stocks drops altogether. As we will use clustering methods based on Euclidean distance and the trading days as features, most stocks can be clustered together.

When the number of stocks contained in a cluster is comparable to or more than the number of observations, the estimation error can remain high and the poorly estimated portfolio weights would bring a negative impact on the portfolio optimization. Therefore, we suppose there is a room for improvement in clustering stocks by the price movement because sometimes the cluster size is so large that the estimation error starts to offset the benefit of diversification coming from portfolio optimization.

The motivation leading our research is that as the clustering size is a matter of issue, we need to explore how to cluster stocks more evenly across clusters by performing data pre-processing and limiting the maximum clustering size. Data pre-processing methods utilized in this research are scaling and dimensionality reductions. Clustering with a constraint on size has been intensively researched in other domains: logistics, urban planning, etc. Our research attempts to cluster stocks for less amount of estimation error but still with reasonable clustering quality.

## 1.2 Research Goal

The goal of our research is to find a price-based clustering algorithm for stocks, which we will use for portfolio optimization to improve the portfolio performance. This clustering algorithm should take care of both estimation error in GMVP and clustering quality because both affects the portfolio performance. The portfolio performance used to evaluate clustering algorithm includes not only risk but also risk adjusted return.

## 1.3 Experiments

Experiments were carried out on 590 U.S. stocks from the Russel 1000 index, which have no missing values from 1999.11.02 to 2019.11.29. The performance of our proposed method is compared to two baseline models, GMVP on individual stocks and GMVP on industry sectors. The stock data is split into validation period and test period so that data belonging to test period would remain unseen until all hyper-parameters such as number of principal components are decided in validation period. The codes for the experiment is also uploaded on Github, so anyone can replicate the experiment described in this paper.[1]

## 1.4 Contributions

The three key contributions of our work are as follows:

- Discover that in clustering approach to portfolio optimization, estimation error and correlation between clusters are trade-off and maximum cluster size influence this relationship. As the maximum cluster size increases, there is less difficulties in clustering similar stocks but higher estimation error while coming up with the covariance matrix of the stocks belonging to the cluster.

---

[1] https://github.com/hellojinwoo/CA_GMVP

- Show that both estimation error and correlation between clusters affect the performance of portfolio optimization and the compromise between these two factors should be found by controlling the maximum cluster size to achieve the best portfolio performance. It implies that we need a clustering algorithm that explicitly allows us to limit the maximum clustering size to explore the connection between the maximum clustering size and portfolio performance.

- Presents that the portfolio performance improvement from scaling data and applying dimensionality reduction methods is practically from reducing the estimation error. It is because the clustering is more balanced when we utilize scaling data or a certain dimensionality reduction method. However, we cannot control the maximum clustering size with these two data pre-processing methods, we still need a clustering algorithm that allows us to manually limit the maximum clustering size.

# 2. Related work

In section 2.1, we present previous researches on global minimum variance optimization. Section 2.2 discusses studies on estimation of covariance matrix used in portfolio optimization, whereas section 2.3 summarizes clustering methods.

## 2.1 Global minimum variance portfolio (GMVP)

Global minimum variance portfolio (GMVP) is an important part of modern portfolio theory (MPT) introduced in 1952 [3]. In the MPT, the return of financial asset is assumed to follow the Gaussian distribution. Therefore, the return of assets is expressed with two statistical properties: mean and variance. As people targets to maximize the return and minimize risk, maximizing mean-variance ratio seems to be a wise objective.

However, it turns out that estimation error in expected returns was 10 times bigger than the estimation error in the variance and covariance [8]. Due to the estimation error, some researchers found that GMVP performed better than model maximizing the risk adjusted return [9]. Furthermore, due to estimation error, the GMVP portfolio weights should be more stable than that of the portfolio optimization utilizing both mean and covariance matrix [10]. To put it differently, small changes in input can cause more drastic changes in portfolio weights if the portfolio optimization depends both on mean and covariance matrix of asset returns.

As such, this dissertation's experiment utilize GMVP for the portfolio optimization. The mathematical formula for the GMVP is provided below.

$$W_{GMV} = \underset{W}{\mathrm{argmin}} \ \{W^T \Sigma W : W^T \cdot 1_N = 1\}$$

$W_{GMV}$ is a portfolio weight vector that we try to obtain by computing the mathematical equation. $\Sigma$ is a covariance matrix of asset returns and $1_N$ is a N dimensional vector of ones.

## 2.2 Estimation of covariance matrix in portfolio

Compared to expected mean of returns, the estimation of variance and covariance parameters of return of assets from historical data is less problematic since the estimation error in variance and covariance is 10 times smaller than estimation error in returns. The smaller estimation error of variance and covariance is due to the fact that the volatility of financial data tends to be non-random and exhibits a positive serial correlation [11, 12]. In other words, we can safely assume that the future variance and covariance parameters will be similar to the historical pattern of the past. As such, using historical price data to come up with the covariance matrix of assets is accepted appropriate.

The sample covariance matrix is an unbiased estimator of population covariance matrix and easy to compute. However, the disadvantage is that the sum of estimation errors contained in each variance and covariance parameter can be so large that it can offset the benefit of diversification [13]. This disadvantage is salient when the number of data points (e.g. daily return or weekly return of assets) is comparable or smaller than the number of individual stocks. The lack of data points is not so uncommon in financial situations because the statistical properties of return of individual stock changes as time goes by. As such, one cannot use stock data covering a long period of time, which causes the lack of data points. When the sample covariance matrix needs estimation of too many parameters with not comparable amount of data, it overfits the sample data because it is of little structure [14]. Moreover, the sample covariance matrix gets closer to singularity as the number of stocks is close to that of observed data. When the covariance matrix is close to singularity, the inverse of it is unstable and sensitive to small change of inputs [15].

One alternative to the sample covariance matrix is the single factor or multi-factor models [16, 17]. It is an estimator with a lot of structure which contain much less estimation error than sample covariance matrix because it has less parameters to estimate. However, it can have more bias and factors can be chosen wrong. To choose the right factors, one should have expert-level domain-specific knowledge to discern important factors from unnecessary factors. Furthermore, in multi-factor models, finding an appropriate number of factors to be included in models has no right answer.

Shrinkage estimators, another alternative to sample covariance matrix, combines sample covariance matrix with a structured estimator to shrink the sample covariance matrix towards the other estimator [18]. The advantage of shrinkage estimator is that we can obtain a 'compromise' between two extreme estimators, a sample covariance matrix and a structured estimator. However, the disadvantage of shrinkage estimators is that even though the true correlation coefficient between two variables is extreme, the shrinkage estimators tend to reduce the extreme value. It is because the shrinkage estimators impose a uniform structure on all covariance value to focus on reducing overestimation of correlation values [19].

Another approach to decrease the estimation error of the sample covariance matrix is clustering similar assets in order to decrease the number of features in a covariance matrix. Clustering method is a structured estimator but has some advantages over other structured estimators. Firstly, unlike a factor model, there is no danger of mis-specifying the factor because it groups similar stocks based on data rather than using a pre-specified predictor. Secondly, unlike shrinkage estimators, clustering stocks do not attempt to reduce extreme values among significantly correlated features.

Rather than coming up with the covariance matrix of all assets all at once, which can contain a great deal of estimation error, clustering approach to covariance matrix is more of 'divide and conquer' method. It splits the stocks into several clusters and comes up with the covariance matrix of each cluster and performs portfolio optimization based on them. With the optimization outcome of each cluster, another portfolio optimization is performed to achieve the portfolio weights of all stocks.

In portfolio optimization, some researches focus on clustering stocks based on non-price information such as accounting figures and industry sectors [20, 21]. However, studies show that the correlation between the accounting figures such as earnings and stock returns has declined in the past 50 years [22-24]. Industry sectors are not good yardsticks to group stocks as well because even if companies belong to the same industry, they can show vastly different price movements in advanced countries [25].

Other researches focus clustering stocks based on the price movement, as the portfolio optimization needs to combine stocks having different price movements to create a stable portfolio. For example, stocks are clustered together if the Pearson correlation coefficient of stock price movements covering certain period is above 0.2 [7]. However, it may lose large part of stock price movement information when the movement is converted to the single number of Pearson correlation coefficient. In another research, K-means clustering is applied on historical daily returns of stocks [26]. However, the clustering algorithm can put too many stocks in one clustering if we do not perform any data pre-processing on stock returns or put the limits on the maximum cluster size. A high dimensionality of covariance matrix results in a high estimation error in portfolio weight vectors even after clustering. As such, in this dissertation we apply data-preprocessing and utilize a bounded clustering algorithm.

## 2.3 Clustering

Clustering refers to a broad set of techniques to find subgroups, or clusters, of unlabeled observations so that the observations within a group are quite similar to each other while different from the observations in other groups. In other words, clustering seeks to partition unlabeled data into distinct groups of high intra-cluster similarity and high inter-cluster dissimilarity.

Clustering problems can be roughly divided into two categories: partitional methods and hierarchical methods [27]. K-means algorithm proposed by MacQueen would be the most famous partitional clustering algorithm [28]. In K-means clustering, we seek to partition the unlabeled data into a pre-specified number of clusters. Even though it is easy to implement in nature, the disadvantage of the algorithm is that the partitions of unlabeled data are highly dependent on initialized points and might not find the global minimum of cost function easily.

On the other hand, in hierarchical clustering, we build nested clusters by merging clusters successively [29]. Unlike K-means clustering, we do not need to pre-specify the number of clusters upfront. However, the computation complexity is higher than K-means clustering and it is not able to undo the previous step.

In some cases, we might have prior information about the underlying cluster structure of the data and would like to guide the clustering process by constraining clusters [30]. For example, one of the drawbacks of K-means clustering is that it may end up local optima and some clusters have zero or very few unlabeled data or too many data. Unbalanced clustering is especially salient when K-means clustering algorithm is used to cluster data having many features or number of clusters is high. To improve upon this drawback, size constraint is added to each cluster of K-means clustering so that a cluster has an upper or lower bounds of number of observations [31, 32].

# 3. Methodological Background

This chapter introduces and reviews clustering algorithms and portfolio performance measures used in this dissertation. Section 3.1 introduces the implementation of clustering algorithms used, whereas section 3.2 explains the several portfolio performance measures that are used to the decide which set of hyper-parameters can create the portfolio with best performance.

## 3.1 Clustering algorithms

We use three clustering algorithms, K-means clustering, hierarchical clustering, and bounded K-means clustering to cluster stocks based on Euclidean distances among a year-long daily return vectors of stocks. As Euclidean distance is practically equal to a distance based on Pearson correlation coefficient, clustering stocks based on Euclidean distance is appropriate to group stocks showing similar daily return movements [33].

Unlike traditional clustering methods that cannot impose constraints on each cluster size, bounded K-means clustering algorithm incorporate size constraints for each cluster separately. Thus, It can be used to obtain clusters in preferred sizes.

The key difference between K-means clustering algorithm and bounded K-means clustering algorithm is assignment of observations to clusters. The former assigns each data point to the closest centroid, while bounded K-means clustering algorithm assigns each data point to the closest centroid only if the clusters belonging to the centroid does not exceed pre-specified cluster size. For instance, if the first closest centroid to a certain point A already creates a cluster whose number of data points is same as the maximum cluster size, then the certain data point A belongs to the second closest centroid. If the second closest centroid also has the maximum number of cluster size, then the certain data point A belongs to the third closest data points, and so on.

## 3.2 Portfolio performance measures

To check which combination of clustering algorithm, dimensionality reduction and scaling method brings the best portfolio performance from several point of view, we use six portfolio performance metrics to measure risk adjusted return as well as risk.

When it comes to risk measures, the most common proxy of risk is standard deviation as we consider any deviation from the expectation as risk. As such standard deviation is natural to be assumed as risk in investment sector. The underlying assumption to use standard deviation as risk is that the stock return follows a Gaussian distribution because volatility can be explained by standard deviation alone in Gaussian distribution.

The downside standard deviation is the concept closely related with standard deviation but only focuses on the return being below the expected return. There have been several literatures arguing that downside standard deviation is better indication of risk than standard deviation. For example, focus group interviews were performed on executive members from eight industries and the conclusion of the interviews was that downside standard deviation is more suitable for risk [34]. It is because standard deviation penalizes both upside and downside volatility equally. However, upside volatility means the increase of stock price, which should be rewarded, not penalized.

The Conditional value at risk (CVaR) at q% level is the expected return, usually loss, on the portfolio in the worst (100-q) % of cases [35]. For example, a one day 95% CVaR of $10 million means that the expected loss of the worst 5% scenarios over a one-day perios is $10 million. In this dissertation, we have used 95% for the level of CVaR, which is the expected return in the worst 5% cases. CVaR is increasingly used in both risk management and portfolio management. For illustration, Basel Committee on Banking Supervision began to use CVaR for calculating market risk capital in the Fundamental Review of the Trading Book (FRTB). CVaR is also known as expected shortfall.

The Maximum Drawdown (MDD) is a measure of the maximum decline from a historical peak in terms of cumulative wealth [36]. As the market tends to drop quickly and moves up slowly, the drawdown comes as a negative surprise to investors. Large MDD usually leads to fund redemptions, so the fund management professionals pay special attention to MDD as the risk measure of choice.

When it comes to risk adjusted returns, two measures are used: Sharpe ratio and Sortino ratio. Sharpe ratio measures the performance of an investment compared to a benchmark after adjusting for its risk [37]. As such, the higher the value is, the better the portfolio optimization result is, because it means the expected or annual return is higher for the same amount of risk.

$$Sharpe\ ratio = \frac{E[R - R_b]}{\sqrt{var[R]}}$$

where R stands for the return of a portfolio while $R_b$ stands for the return of benchmark.

Sortino ratio measures the risk-adjusted return as the Sharpe ratio does, but penalizes only those returns falling below a user-specified target [38]. The drawback of the Sharpe ratio is that it penalizes both upward volatility and downward volatility. Along the same line with the Sharpe ratio, the higher the Sortino ratio is, the better the portfolio optimization result is.

$$Sortino\ ratio = \frac{E[R - R_b]}{\delta_d}$$

where R stands for the return of portfolio, $R_b$ stands for the return of benchmark and $\delta_d$ is a standard deviation of returns below the user-specified target.

# 4. Experiment Methods

This chapter presents the design of experiment employed in this research in detail. Section 4.1 introduces dataset used for the experiment, whereas section 4.2 discusses the experiment design.

## 4.1 Dataset

We use stock prices of companies listed on the Russel 1000 index, which is known to be an efficient representation of the overall U.S. stock market. The stock price is used only if the companies does not have missing values of the stock price for the consecutive 20 years, from November $2^{nd}$, 1999 to November $29^{th}$, 2019. 590 companies in the Russel 1000 index as of November 2019 satisfied the criteria. The table 1 presented below shows the number of companies belonging to each of 11 Global Industry Classification Standards (GICS) sectors.

As the table shows, the industrial composition of the stocks used for the thesis has little difference to that of S&P 500, which is another index to be known for representing the overall U.S. stock market. The pie chart of industry composition of stocks used for the experiment and S&P 500 is also provided in the figure 1 below. The left pie chart shows the share of industries that is used for the experiment, while the right pie chart shows the share of industries in S&P 500 as of November 2019.

|  | Experiment data | S&P 500 |
|---|---|---|
| Communication | 18 | 26 |
| Consumer discretionary | 73 | 64 |
| Consumer staples | 37 | 33 |
| Energy | 32 | 28 |
| Financials | 92 | 67 |
| Health care | 68 | 60 |
| Industrials | 93 | 70 |
| Information Technology | 67 | 69 |
| Materials | 30 | 28 |
| Real estate | 49 | 32 |
| Utilities | 31 | 28 |
| sum | 590 | 505 |

**Table 1. Comparison of Industry composition**

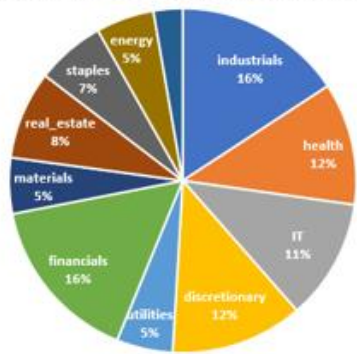
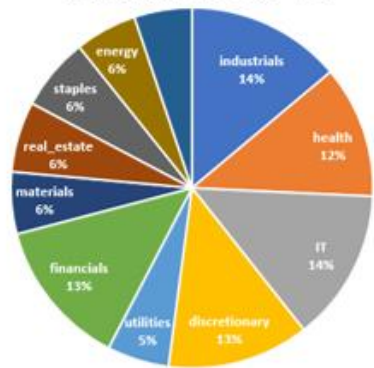

**Figure 1. Industry composition pie chart**

Stock prices used for the experiment were gathered from *Yahoo! Finance* website using a python library *pandas-datareader*. This research uses adjusted closing prices shown in *Yahoo! Finance*, which accounts for stock splits and dividend. It allows investors a fair comparison among stock prices regardless of characteristics of companies. For example, even though a market share of a certain company A is higher than that of the company B, stock price of the company A can be lower than the company B because of the number of issued stocks. To prevent from unfair comparison of stock prices, this experiment uses adjusted closing price.

## 4.2 Experiment design

**Objective of experiment**

The objective of this dissertation's experiment is presented in figure 2.

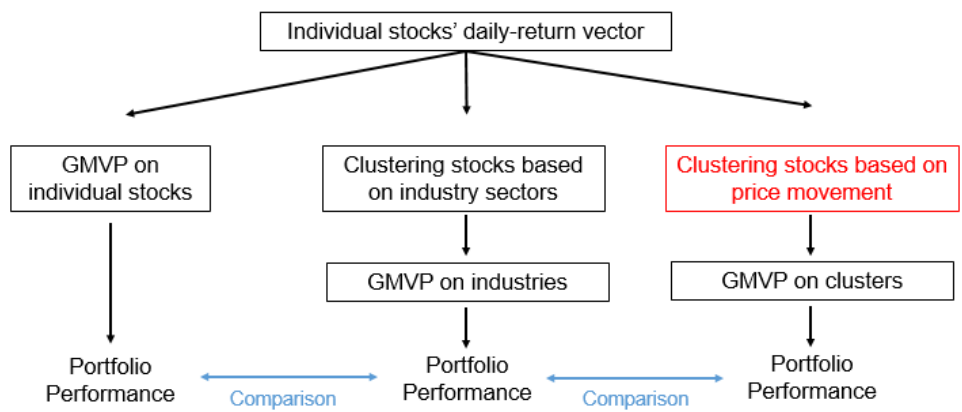

**Figure 2. Objective of experiment**

As the research goal is to find the price-based clustering method for improving the performance of GMVP, the objective of the experiment is finding the price-based clustering algorithm with the best portfolio performance and comparing the performance against benchmarks: GMVP with individual stocks and GMVP with clusters based on non-price information. The foremost performance to be compared against other models is the standard deviation, as the cost function that GMVP attempts to decrease is the standard deviation of portfolio returns. And then other portfolio performance measures such as risk adjusted return and other risk measures, introduced in the section 3.2, will be compared as well

## 4.3 Experiment procedure

To begin with, we use a year-long stock daily returns because the daily returns of stocks show non-stationarity [39]. It means that statistical properties such as mean and covariance in the financial market is continuously and significantly changing over time. For instance, it is unrealistic to assume that the covariance of two certain stocks will remain the same or at least similar for long period of time. As it is proven that covariance remains similar for relatively long period of time, we use one-year long data to come up with the covariance matrix of stocks. The flowchart of experiment procedure is presented in figure 3.

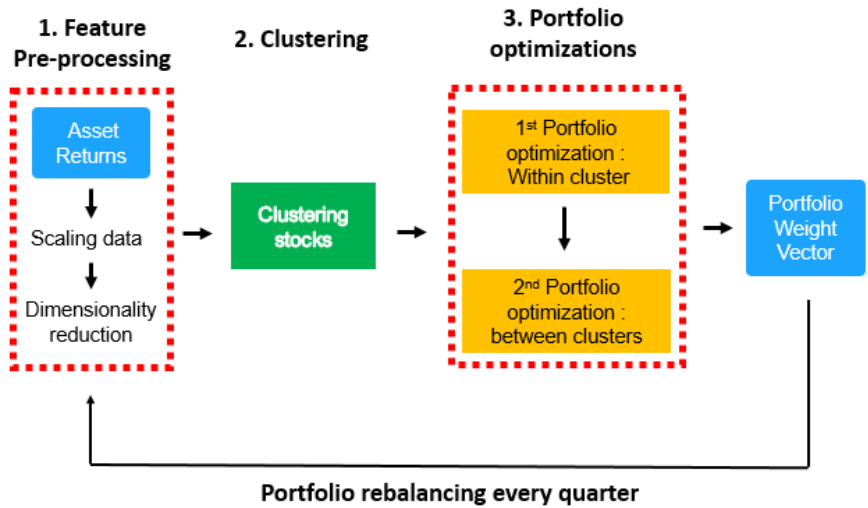

**Figure 3. Flowchart of experiment procedure**

**1st step. Pre-processing daily return**

The first step of the experiment is to pre-process the daily return of stocks before clustering. We also experiment with data that does not go through pre-processing to experiment which methods produce the best portfolio performance. Two data pre-processing methods utilized in this experiment is scaling and dimensionality reduction.

Standard scaling is utilized to make all the features contribute equally while clustering stocks. As we use Euclidean distance in clustering stocks in this dissertation, features with larger variance affects the Euclidean distance measured in each dimension. The features with more variance will have more difference of Euclidean distance and thus will be over-represented while clustering stocks. Since standard scaling makes all the features have the same unit variance, the difference of importance in features is removed and all the trading days are equally represented. For instance, when the stock market crashes, stock price of some stocks drops by more than 10% while price of some stocks stay almost the same, resulting in a feature with large variance. It leads to the difference in Euclidean distance in this dimension of features, so larger weights in clustering stocks.

Another pre-processing applied on data is dimensionality reduction method. One-year long trading day vector is 252-dimensional vector so it might be considered high-dimensional, which causes 'curse of dimensionality' [40]. In high-dimensional space, Euclidean distance loses its meaning and does not work well so clustering may fail [41]. To be more specific, the concept of distance becomes less precise as the number of data features grow, since the distance between any two points in a given dataset converges. Furthermore, if the number of data features approach infinity, all the distances between observations would be equal. The discrimination of the nearest and farthest point for clustering data becomes impossible. Dimensionality reduction decreases the dimensionality of feature space, so curse of dimensionality can be prevented. Dimensionality reduction is also known to decrease the noise in data, which yields better classification and clustering results [42]

Two dimensionality reduction is applied respectively: Principal Component Analysis (PCA) and t-Stochastic Neighbor Embedding (t-sne). PCA creates a linear combination of features to explain as much variance as possible, because it assumes that features with more variance is more important to explain data. T-sne is also a feature extraction method but creates features by non-linear combinations of features. The appropriate number of principal components and t-sne components need to be found through grid-search to achieve the best portfolio performance.

**2nd step. Clustering stocks & 'within-cluster' portfolio optimization**

With the pre-processed or raw-data, we cluster stocks by pre-specified number of groups. In this experiment, we cluster stocks by 11 groups for the purpose of comparison with results of the previous research, experimented with 11 industry sectors classified by GICS. The industry classification of stocks used in this experiment is shown in the section 4.1.

After clustering, the GMV portfolio optimization is performed within each and every cluster. The stock universe is limited to the stocks within a cluster, so it creates a covariance matrix with the decreased number of features while performing GMV optimization. This decreased number of features decrease the estimation error in the estimated covariance matrix. After the portfolio weights of stocks within a cluster is obtained, we construct a portfolio and consider the portfolio as a single stock. The portfolio, an outcome of 'within-portfolio' optimization, is considered a stock that can be traded in the stock market. With these 11 portfolios, we continue to the second stage of portfolio optimization, which is 'between-clusters' optimization.

**3rd step. 'Between-clusters' portfolio optimization**

'Between-clusters' optimization is not different from regular GMVP optimization with stocks, except for that the GMVP optimization is performed on portfolios. The outcome of the 'between-clusters' optimization is the portfolio weights, indicating how much weight of wealth should be invested in each cluster to achieve the least amount of volatility.

With the outcome of 'between-clusters' GMVP optimization and 'within-cluster' optimization, we can figure out how much weight of wealth should be invested in each stock. Then following the portfolio weights of each stock, we invest accordingly for 3 months. After 3 months, we repeat this process to rebalance the portfolio weights.

**Portfolio rebalancing**

The diagram of how to rebalance portfolio is shown in the figure 4.

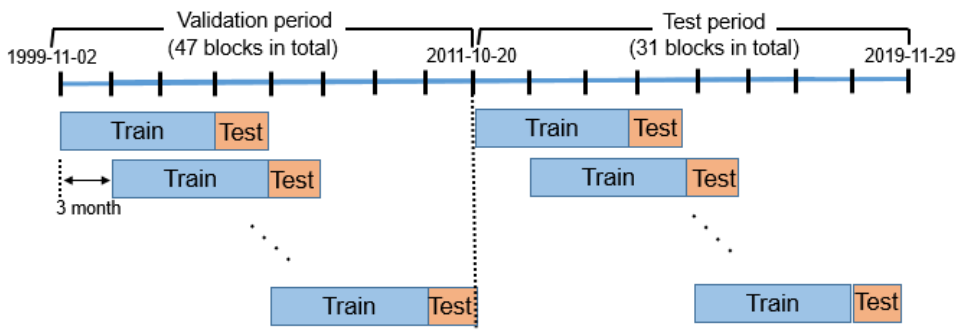

**Figure 4. Portfolio rebalancing diagram**

The portfolio optimization is executed 47 times in the validation period. During the validation period, several combinations of scaling method, dimensionality reduction method, clustering method, and other hyper-parameters such as learning rate and perplexity in t-sne are experimented. Then with the combinations showing the least amount of volatility (standard deviation), we apply them in the test period to compute the test performance. We rebalance the portfolio 31 times during the test period and compare the performance of each method. The performance used for comparing models are not only standard deviation but also risk adjusted returns and other forms of risk such as maximum drawdown. The result of performance is provided in tables in section 5.1 and section 5.2.

# 5. RESULTS

In this chapter, we present the outcome of our experiment. In section 5.1, we show standard deviations of portfolios created with different combinations of methods handling data. In section 5.2, estimation error of benchmarks and the model with best performance is compared, whereas several portfolios performance are compared in section 5.3.

## 5.1. Model Evaluation

Clustering models are evaluated using standard deviation of daily return of a portfolio during the out-of-sample period (test period). However, the standard deviation during the in-sample period (validation period) is also presented along the out-of-sample period performance. It is because we selected the hyper-parameters of models such as number of principal components in PCA or perplexity and learning rate in t-sne by the performance of the in-sample period.

Performance of portfolios is reported using only standard deviation while comparing different clustering algorithms because the objective function of the GMVP is to minimize the standard deviation. GMVP assumes that return of stocks follow a Gaussian distribution, so all the risk, deviation from the expectation, can be explained by the standard deviation alone. In the section, 5.3, the other portfolio performance measures such as risk adjusted return and other forms of risk will be presented.

The comparison table showing the result of experiment, each model's standard deviation, is shown below in table 2. We experimented for 100 times with different combination of several hyper-parameters, and computes its mean of standard deviations.

| Clustering | Dim. reduction | Scaling method | Validation std | Test std |
|---|---|---|---|---|
| GMVP on individual stocks | | | 0.1068 | 0.0951 |
| GMVP on industry sectors | | | 0.0948 | 0.0845 |
| K-means Clustering | Not used | Standard Scaled | 0.1009 | 0.0954 |
| | | Raw data | 0.1902 | 0.2053 |
| | PCA | Standard Scaled | 0.0989 | 0.0911 |
| | | Raw data | 0.2205 | 0.1916 |
| | t-sne | Standard Scaled | 0.0935 | 0.0825 |
| | | Raw data | 0.0967 | 0.0829 |
| Hierarchical Clustering | Not used | Standard Scaled | 0.1022 | 0.0943 |
| | | Raw data | 0.1075 | 0.0948 |
| | PCA | Standard Scaled | 0.1038 | 0.0946 |
| | | Raw data | 0.1070 | 0.0949 |
| | t-sne | Standard Scaled | 0.0952 | 0.0843 |
| | | Raw data | 0.0962 | 0.0864 |
| Bounded K-means Clustering | Not used | Standard Scaled | 0.0906 | 0.0825 |
| | | Raw data | **0.0886** | **0.0798** |
| | PCA | Standard Scaled | 0.0906 | 0.0822 |
| | | Raw data | 0.0900 | 0.0805 |
| | t-sne | Standard Scaled | 0.0925 | 0.0872 |
| | | Raw data | 0.0905 | 0.0862 |

**Table 2. Standard deviation of portfolio daily returns (annualized)**

As one can see from the table 2, the performance of GMVP with bounded K-means clustering on raw-data without using dimensionality reduction shows the least amount of standard deviation. Compared to the two baseline models, GMVP on individual stocks and GMVP on industry sectors, the proposed method improves the portfolio performance by decreasing the risk by 15.6% and 5.6% respectively

There are two interesting things to notice in the table 2. Firstly, the standard deviations of GMVP with raw-data are generally higher than GMVP with standard scaled data, especially in K-means clustering. The standard deviations of GMVP with K-means clustering on raw data without dimensionality reduction or PCA is two times bigger than that of GMVP with K-means clustering on standard scaled data. GMVP with hierarchical clustering also shows poor performance with raw data, depending on linkage methods. For example, if the linkage method 'ward' is used with raw data without the dimensionality reduction method, it shows the annualized standard deviation of 0.1918.

Secondly, when the stocks are clustered with data embedded through t-sne, the performance improves in general except for the bounded K-means clustering. These two findings are related with the trade-off between correlation between clusters and estimation error in the covariance matrix of clusters. The reason will be discussed in more detail in the next chapter, 6. discussion, showing how dimensionality reduction and scaling affects the standard deviation of portfolio returns,

## 5.2 Estimation error

|  | In-sample Std | Out-of-sample Std | Estimation error |
|---|---|---|---|
| stock-based GMVP | 0.0489 | 0.0946 | 93.51% |
| Industry-based GMVP | 0.046 | 0.0846 | 83.79% |
| Cluster-based GMVP | **0.0462** | **0.0798** | **72.73%** |

**Table 3. Estimation error of portfolios (annualized)**

The difference between 'In-sample standard deviation' and 'Out-of-sample standard deviation' is presented above, in table 3. The difference between the in-sample Std and Out-of-sample Std is defined as 'estimation error' because we expect the variance and covariance of stocks would remain similar for about one year and the statistical properties of out-of-sample period is estimated using the in-sample period counterpart. The in-sample standard deviation is the mean of standard deviation of portfolio daily returns for one year. Out-of-sample standard deviation is the mean of daily return of portfolio for three months.

The high estimation error means that the sum of errors in the estimated covariance matrix contains high degree of error. Moreover, the inverse matrix of the covariance matrix is unstable and sensitive, so small error in the covariance matrix gets amplified and will lead to a very different inverse. The benefits of diversification can be more than offset by estimation errors, so GMVP can underperform naïve asset allocation such as 1/N allocation. With the less percentage of difference, we can confirm that the proposed method has less estimation error than baseline models

## 5.3. Several performance measures

|  | Sharpe Ratio | Sortino Ratio | Std | Downside Std | Maximum DrawDown | CVaR |
|---|---|---|---|---|---|---|
| stock-based GMVP | 0.8963 | 1.2915 | 0.0946 | 0.0686 | -15.69% | -1.12% |
| Industry-based GMVP | 1.8232 | 2.5207 | 0.0848 | 0.0637 | -9.36% | -0.97% |
| Cluster-based GMVP | **1.8316** | **2.5726** | **0.0803** | **0.0608** | **-8.21%** | **-0.93%** |

**Table 4. Several portfolio performance measures (annualized)**

One experiment result of bounded K-means clustering results is provided using several portfolio performance measures in the table 4 above. As we can see from the table, all the portfolio performance measures improve compared to baseline models. Although we do not have large difference in the two measures of risk adjusted return, Sharpe ratio and Sortino ratio, we can relatively great improvement in risk performances: 5.3% in standard deviation, 4.6% in downside standard deviation, 12.3% in Maximum Drawdown and 4.1% in Conditional Value at risk.

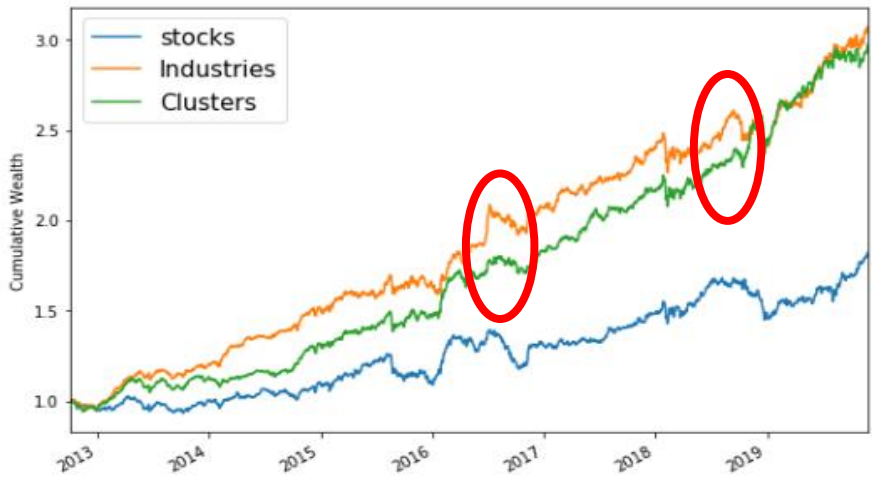
**Figure 5. Cumulative wealth line graph**

In terms of the cumulative wealth, we cannot see a big difference between industry-based GMVP and cluster-based GMVP. The figure 5 above shows the cumulative wealth from the end of 2012 to the end of 2019, for the consecutive 7 years. The smaller standard deviation can be confirmed in this cumulative wealth line graph as well. For example, in the middle of the year 2016 and 2018, the green line graph shows less volatile movement than the yellow line graph. The green line graph and yellow line graph represent the cluster-based GMVP and industry-based GMVP, respectively.

# 6. Discussion

To put bottom line up front, we found three points in role of maximum cluster size in clustering approach to portfolio optimization. Firstly, there is a trade-off between estimation error and clustering quality while we control the maximum cluster size. Both of these affect the portfolio performance so controlling the maximum cluster size influence portfolio performance. Secondly, due to the trade-off relationship, we need to find where to set the maximum clustering size for figuring out where to compromise between the clustering quality and estimation error. Thirdly, portfolio performance improvement from data scaling and dimensionality reduction is also deeply related with the trade-off between clustering quality and estimation error. It is because scaling data and dimensionality reduction helps creating balanced clusters, resulting in decreasing the estimation error.

## 6.1 Trade-off between correlation and estimation error

By constraining the maximum size of clusters, it is unavoidable for the quality of clusters to deteriorate because the constraint hinders the objective of clustering: high intra-cluster similarity and high inter-cluster dissimilarity. Even though some elements show similar set of features, they might end up belonging to a different cluster due to the clustering size constraint. Since similar observations fall into different clusters, inter-cluster dissimilarity decreases.

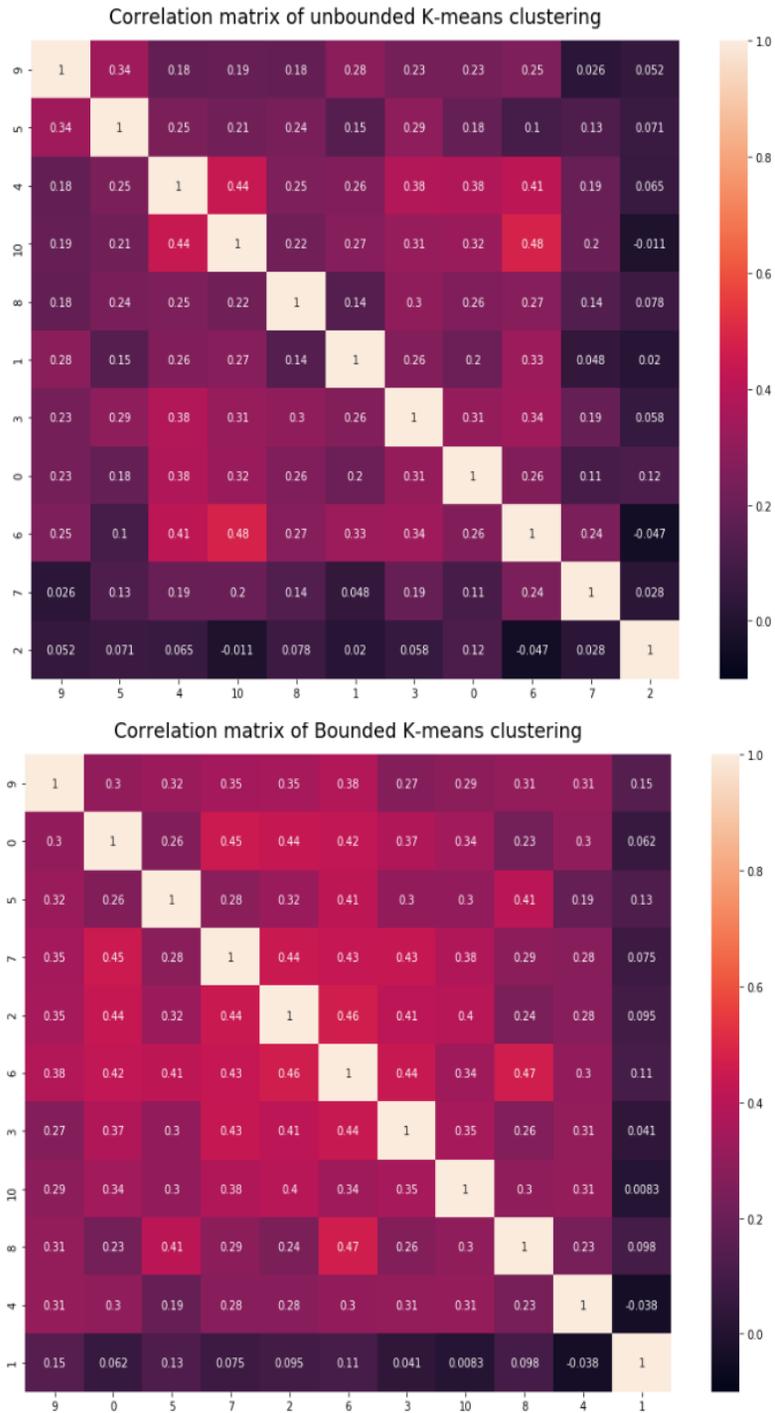

**Figure 6. Heatmaps of Correlation between clusters**

**(upper : ordinary K-means clustering,
lower : bounded K-means clustering)**

Furthermore, maximum clustering size constraint increases correlation between portfolio returns of 'within-cluster' optimized portfolios. As stocks with similar price movements belong to different clusters due to the size constraint, portfolio optimizations within clusters are performed on similar stocks, resulting in the portfolios with similar price movement as a consequence. We can confirm the increased correlation between clusters in heatmaps of ordinary K-means clustering and bounded K-means clustering. The heatmaps are provided in the figure 6 above.

The correlation between clusters created with ordinary K-means clustering is lower than that of bounded K-means clustering. The two correlation heatmaps provided in the figure 6 is created from the daily return of 590 stocks used in the experiment for 1 year, from the beginning of 2012 to the end of 2012. The upper correlation heatmap shows correlation of clusters created with the K-means clustering on raw data without using the dimensionality reduction with the least standard deviation in the table 2. The lower correlation heatmap shows correlation of clusters created with bounded K-means clustering on raw data without using the dimensionality reduction methods, whose maximum clustering size is 75. The degree of correlation is represented with color in the heatmap. The brighter the colors are, the more correlated between random variables are. The range of correlation expressed in the plot is from 1 to -0.1, because no correlation value is smaller than -0.1 in this example.

However, advantage of having smaller number of maximum cluster size is the smaller estimation error contained in covariance matrix and inverse of covariance matrix. With the reduced number of stocks, the dimensionality of covariance matrix becomes smaller. If the amount of observations is the same, reduced number of features results in less estimation error in GMVP optimization process because there are less number of variance and covariance parameters to be estimated and the covariance matrix gets farther away from singularity.

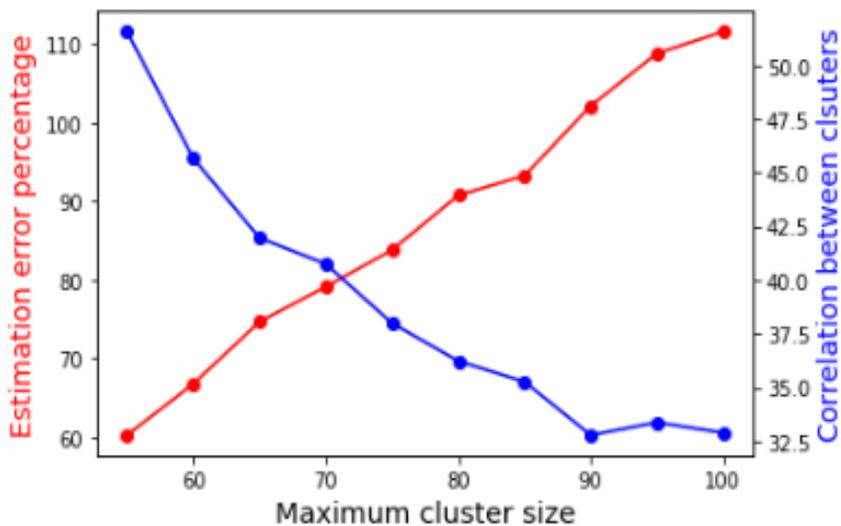

**Figure 7. Trade-off due to maximum cluster size**

The trade-off line graph of 'correlation between clusters' and 'estimation error percentage' is provided in the figure 7 above. The estimation error and correlation between clusters are measured with bounded K-means clustering on raw-data without using dimensionality reduction methods. We can see in the figure 7 that as the maximum cluster size increases, the estimation error increases while the correlation between clusters decreases.

## 6.2 Finding the compromise for the best performance

As the estimation error and correlation between clusters establish a trade-off and both of them affects the portfolio performance, we should find where to set the maximum cluster size in order to achieve the best portfolio performance. It is because as shown in the previous section 6.1, clustering quality and estimation error is decided by maximum clustering size and both of them are major factors in determining the portfolio performance.

The maximum clustering size plays an important role to minimize the risk in a clustering approach to GMVP because usually the ordinary clustering algorithm focuses only on increasing cluster quality and thus fails in taking care of estimation error. To put it differently, it focuses only on the minimizing the correlation between clusters, not taking care of estimation error. As such, the GMVP with ordinary clustering algorithm ends up not achieving the results as great as the bounded K-means clustering algorithm.

However, if the maximum clustering size is relatively small, clustering quality is poor and thus, correlation between clustering is high. However, the estimation error would be small. On one hand, the poor cluster quality has a negative impact on portfolio performance as more uncorrelated the assets, the less variance the portfolio returns have. On the other hand, the smaller estimation error has a positive impact on portfolio performance. It has a positive impact on portfolio optimization because the outcome of GMVP formula, which is the portfolio weight vector that we invest accordingly, is more stable and has less estimation error. As such, the result of portfolio optimization process gets more accurate, resulting in better portfolio performance.

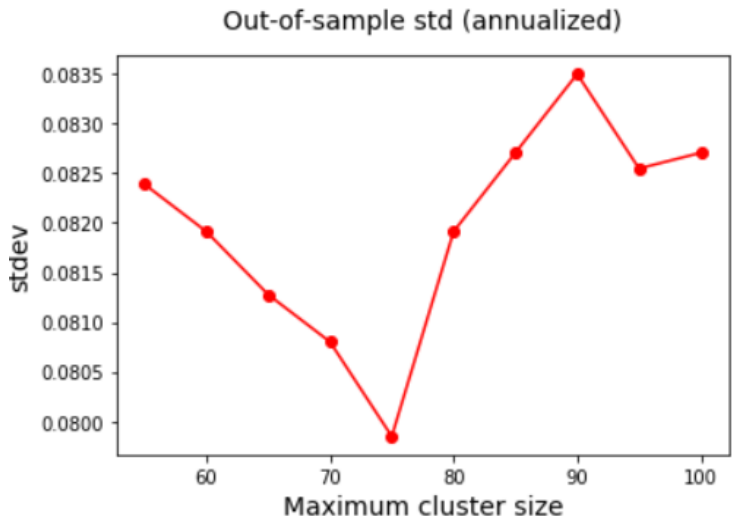

**Figure 8. Relationship between cluster size and performance**

In the figure 8 shown above, the relationship between maximum cluster size and the out-of-sample standard deviation is illustrated as a line graph. The standard deviation of portfolio return is computed from the result of GMVP with bounded K-means clustering on raw data, without using dimensionality reduction method. The V shaped line graph implies that to achieve the best performance of GMVP, we should pay attention to both of correlation between clusters and estimation error due to trade-off. If we take care of only one effect, then the GMVP ends up achieving not the best performance

## 6.3 Effect of feature pre-processing

**Standard scaling**,

We can see that the performance of GMVP with K-means clustering and hierarchical clustering on standard scaled data show lower standard deviation of daily returns of a portfolio, which is better portfolio performance, than raw data consistently. The table 5 shown below is the excerpt from the table2, and we can confirm that the difference between standard deviations is especially large in GMVP with K-means clustering on data, whose dimensions are reduced by PCA, and not reduced by a dimensionality reduction method. The difference is so large that the standard deviations of GMVP with clustering methods on raw data is almost two times bigger than that of GMVP with clustering methods on standard scaled data.

| Clustering | Dim. reduction | Scaling method | Validation std | Test std |
|---|---|---|---|---|
| K-means Clustering | Not used | Standard Scaled | 0.1009 | 0.0954 |
| | | Raw data | 0.1902 | 0.2053 |
| | PCA | Standard Scaled | 0.0989 | 0.0911 |
| | | Raw data | 0.2205 | 0.1916 |
| | t-sne | Standard Scaled | 0.0935 | 0.0825 |
| | | Raw data | 0.0967 | 0.0829 |
| Hierarchical Clustering | Not used | Standard Scaled | 0.1022 | 0.0943 |
| | | Raw data | 0.1075 | 0.0948 |
| | PCA | Standard Scaled | 0.1038 | 0.0946 |
| | | Raw data | 0.1070 | 0.0949 |
| | t-sne | Standard Scaled | 0.0952 | 0.0843 |
| | | Raw data | 0.0962 | 0.0864 |

**Table 5. Standard deviations of clustering algorithms**

The performance improvement due to standard scaling data can be explained from the perspective of decreased estimation error due to more balanced clusters, resulting in smaller dimensionality of covariance matrix of each cluster. Since the standard scaling on features make every feature have the unit variance, all the features contribute equally while clustering stocks. It is because the clustering algorithms utilized in this experiment are based on Euclidean distance of observations. As such, the features with bigger variance would have bigger difference of Euclidean distance and hence, bigger weights of importance while clustering stocks. after standard scaling data, clustering algorithms based on Euclidean distance consider all the features equally because all the features are of the same variance, which is the unit variance.

The outcome of applying standard scaling is more balanced clusters with less concentration of observations in few clusters. For example, outcome of clustering stocks by K-means clustering with raw features and standard scaled features for one year in 2012 is provided below in table 6, for the purpose of comparison. The observations are more concentrated in few clusters when we cluster stocks using raw-data. It means that some trading days which shows bigger difference in stock price play a critical role in deciding which stocks should be clustered together. It is not so uncommon in stock market when most of stocks move towards the same direction due to the systematic risk, also known as market risk. For example, the change of interest rates and news about recessions and wars can make most of stocks can go towards the certain direction. As such, when all the features are standard scaled to have the same unit variance, the observations are now more clustered evenly across clusters.

|  | Standard_scaled | Unscaled |
|---|---|---|
| **cluster_1** | 20.85% | 34.92% |
| **cluster_2** | 15.76% | 20.17% |
| **cluster_3** | 10.85% | 18.81% |
| **cluster_4** | 10.17% | 7.80% |
| **cluster_5** | 8.81% | 7.12% |
| **cluster_6** | 7.63% | 6.95% |
| **cluster_7** | 7.12% | 2.20% |
| **cluster_8** | 5.25% | 1.53% |
| **cluster_9** | 5.25% | 0.17% |
| **cluster_10** | 4.41% | 0.17% |
| **cluster_11** | 3.90% | 0.17% |
| **Sum** | 100% | 100% |

**Table 6. Share percentage of clusters - Scaling**

We can observe these balanced clusters throughout the test period when we use the standard scaled data while clustering stocks. On average, about 88 percent of all observations belong to the top 5 clusters when we perform K-means clustering on raw-data without using dimensionality reduction methods, whereas only about 65 percent of all observations in top 5 clusters when we perform the same clustering method, K-means clustering, on standard scaled data. The line graph of share percentage of clusters using different scaling method is shown in the figure 9 below.

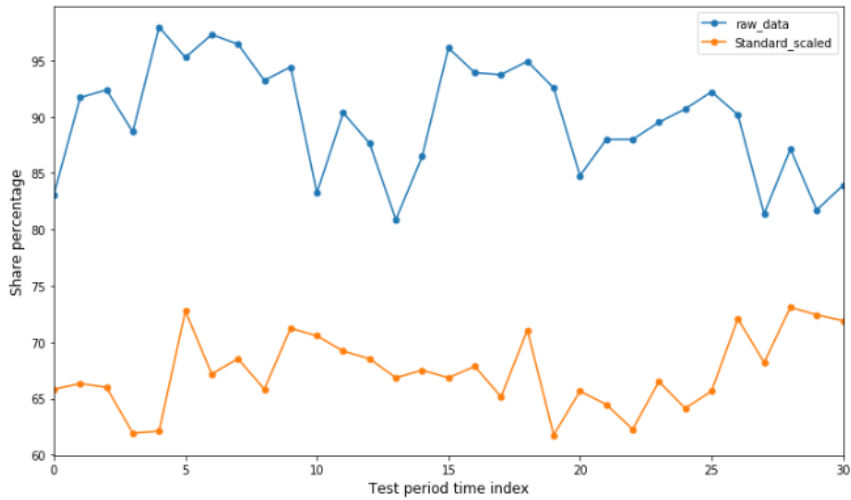

**Figure 9. Share percentages of top 5 clusters - Scaling**

As we saw from the figure 8, only when we take care of both clustering quality and estimation error, the portfolio performance improves. Standard scaling has the impact of decreasing estimation error by clustering stocks more evenly, as we can confirm in the figure 9. The blue line shows the share of top 5 clusters of K-means clustering on raw data expressed in percentage. The orange line shows the share of top 5 clusters of K-means clustering on standard scaled data.

However, standard scaling is not enough to achieve the best portfolio performance because we cannot precisely control the maximum clustering size. The cluster size is still decided by the clustering algorithm to maximize the clustering quality, so the cluster size is most likely not optimized for minimizing the risk. As such, it implies that we need a clustering algorithm that allows users to control the maximum cluster size.

**Dimensionality reduction methods**

From the table 5, we can also confirm that clustering stocks based on t-sne greatly improves the portfolio performance, compared to clustering stocks based on raw-data or data embedded through PCA. Regardless of whether the data is standard scaled or not, clustering stocks with data embedded with t-sne shows lower standard deviation of daily returns of portfolio. We can also explain this from the perspective of trade-off between estimation error and clustering quality.

When stocks are clustered with features embedded through t-sne, they are much less concentrated in few clusters than clustering stocks with data embedded through PCA and data without using dimensionality reduction methods. The outcome of clustering stocks by K-means clustering with dimensionality reduction methods on un-scaled data for one year in 2012 is provided below in table 7. The results of table 7 illustrates that clustering data with t-sne embedded features can create more balanced clusters. The balanced clusters created with t-sne is consistent throughout the test period, as we can see in the figure 10. The line graph in figure 10 shows the top 5 clusters' share expressed in percentage. The blue line is clustered with data without using a dimensionality reduction method, the orange line shows share of each clusters with data embedded through PCA, and the green line shows the clusters grouped with data embedded through t-sne. As such, the portfolio performance improvement from clustering stocks with embedded by t-sne comes from decreasing the estimation error, because the dimensionality of covariance matrix of a certain cluster is not too large.

|  | t-sne | PCA | Not dim. reduced |
|---|---|---|---|
| **cluster_1** | 11.00% | 49.92% | 45.69% |
| **cluster_2** | 10.49% | 21.49% | 19.63% |
| **cluster_3** | 10.32% | 11.68% | 19.29% |
| **cluster_4** | 9.48% | 8.97% | 8.12% |
| **cluster_5** | 9.14% | 3.21% | 3.38% |
| **cluster_6** | 8.63% | 1.69% | 2.54% |
| **cluster_7** | 8.63% | 1.52% | 0.51% |
| **cluster_8** | 8.46% | 0.85% | 0.17% |
| **cluster_9** | 8.46% | 0.17% | 0.17% |
| **cluster_10** | 7.78% | 0.17% | 0.17% |
| **cluster_11** | 7.45% | 0.17% | 0.17% |
| **Sum** | 100% | 100% | 100% |

**Table 7. Share percentage of clusters - Dim. reduction**

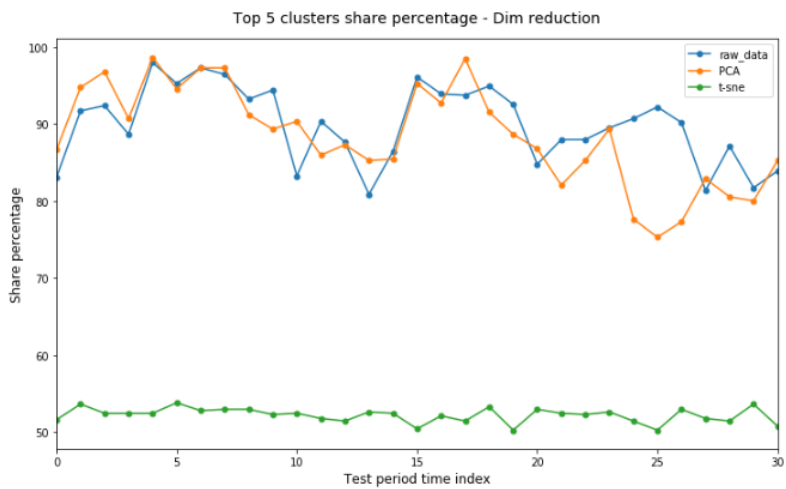

**Figure 10. Share percentages of top 5 clusters - Dim. reduction**

# 7. Conclusion

Due to the recent market crashes, investors participating in stock market has a great deal of interest in investing their wealth safely. GMVP is appropriate investment strategy for addressing the needs of safe investment because it is designed to carry as little variance as possible. Since it needs only covarinace matrix of sotkcs that are being considered for invesmtnet as its input to compute the portfolio weights, it has less estimation error than other portfolio optimization methods utilizing both mean and covaraince matrix. However, if the number of stocks is comparable to or bigger than that of observations, the covariance matrix contains great deal of estimation error and inverse matrix is unstable. Some researchers argue that equally weighted portfolio, which does not go through portfolio optimization, performs better than GMVP because the benefit of diversification is more than offset by the estimation error and instability of inverse covariance matrix.

To reduce the estimation error in the covariance matrix and avoid approaching singularity, several attemts have been made and one of them is clustering approach. The bottom line of the clustering approach is 'divide and conquer'; because it is difficult to estimate the covariance matrix of all the stocks all at once when the number of stocks is high, it divides the stocks into several clusters and performs portfolio optimization in each cluster first. With the outcome of portfolio optimization in each cluster, another portfolio optimization is performed once again between clusters.

Many clustring methods have been suggested for clustering stocks but they can be roughly divided into two categoris: clustering based on non-price information such as accounting figures and clustering based on price. Since non-price information often diverges from the direction of price, the clustering quality becomes poor. More uncorrelated clusters are, the less variance of portfolio returns is. As such, clustering based on price is more appropriate for the portfolio optimization. However, the problem of high estimation error and unstable inverse of covarinace matrix may still remain if too many stocks are clustered in a single group. In certain period of trading days, It leads to the high dimensionality of covariance matrix which results in high estimation error while performing portfolio optimization within each cluster.

Therefore, in this dissertation, we explore how to cluster stocks based on price, focusing on not only decreasing the estimation error but also increasing the clustering quality. Three clustering algorithms are utilized: K-means clustering, hierarchical clustering, and bounded K-means clusteirng algorithm. Two baseline models, GMVP based on individual stocks and GMVP based on industry sectors are used for comparison.

With the bounded K-means clustering algorithm, we reduced the out-of-sample standard deviation of portfolio returns by 15.7% and 5.6% than the baseline models, respectively. Furthermore, the gap between the out-of-sample and in-sample counterpart also decreases, so we can estimate the out-of-sample risk with better accuracy. Other porfolio performance measures such as risk adjusted return and maximum drawdowns also improves with bounded K-means clustering.

While experimenting, we learned three critical implications for people considering a clustering approach to GMVP. Firstly, there is a trade-off between correlation between clusters and estimation error and maximum clustering size can control this trade-off. As the maximum clustering size increases, the correlation between clusters decreases, whereas the estimation error gets larger and vice versa. Secondly, because we can manually set the maximum clustering size with bounded clustering algorithms, the level of estimation error and clustering quality can be controlled. As both of these determine the portfolio performance, we can find the compromise point for the best portfolio performance. Thirdly, scaling data or utilizing dimensionlaity reduction on data can also reduce portfolio risk and it can be interepreted as decreasing estimation error. However, unlike bounded K-means clustering, we cannot precisely control the maximum clustering size, so scaling and dimensionality reduction methods may fail to achieve the best portfolio performance. This implies that we need a clustering algorithm that allows us to control the maximum clustering size.